\documentstyle[12pt]{article}

\def\hybrid{\topmargin -20pt  \oddsidemargin 0pt
      \headheight 0pt   \headsep 0pt
      \textwidth 6.25in 
      \textheight 9.5in 
      \marginparwidth .875in
      \parskip 5pt plus 1pt   \jot = 1.5ex}

\hybrid

\begin{document}
\def\x{\times}
\def\beq{\begin{equation}}
\def\eeq{\end{equation}}
\def\beqa{\begin{eqnarray}}
\def\eeqa{\end{eqnarray}}

\sloppy
\newcommand{\be}{\begin{equation}}
\newcommand{\eq}{\end{equation}}
\newcommand{\ov}{\overline}
\newcommand{\un}{\underline}
\newcommand{\p}{\partial}
\newcommand{\la}{\langle}
\newcommand{\ra}{\rangle}
\newcommand{\bl}{\boldmath}
\newcommand{\ds}{\displaystyle}
\newcommand{\nl}{\newline}
\newcommand{\th}{\theta}



\renewcommand{\thesection}{\arabic{section}}
\renewcommand{\theequation}{\thesection.\arabic{equation}}

\parindent1em


\begin{titlepage}
\begin{center}
\hfill HUB-EP-97/36\\
\hfill {\tt hep-th/9706093}\\

\vskip .7in

{\bf Three-Branes and Five-Branes in $N=1$ Dual String Pairs}

\vskip .3in

Bj\"orn Andreas and Gottfried Curio\footnote{email: 
andreas@qft3.physik.hu-berlin.de, curio@qft2.physik.hu-berlin.de}
\\
\vskip 1.2cm

{\em Humboldt-Universit\"at zu Berlin,
Institut f\"ur Physik, 
D-10115 Berlin, Germany}

\vskip .1in

\end{center}

\vskip .2in

\begin{center} \end{center}
\begin{quotation}\noindent

In this note we show that in dual $N=1$ string vacua provided by the heterotic
string on an elliptic Calabi-Yau together with a vector bundle respectively
$F$-theory on Calabi-Yau fourfold the number of heterotic fivebranes
necessary for anomaly cancellation matches the number of $F$-theory
threebranes necessary for tadpole cancellation. This extends to the general 
case the work of Friedman, Morgan and Witten, who treated the case of 
embedding a heterotic $E_8\times E_8$ bundle, leaving no unbroken gauge group,
where one has a smooth Weierstrass model on the $F$-theory side.

\end{quotation}
\end{titlepage}
\vfill
\eject

\newpage

 In the course of study of $F$-theory compactifications with $N=1$
supersymmetry in four dimensions on a Calabi-Yau fourfold $X$
there was established the necessity
of turning on a number $n_3=\chi (X)/24$ of spacetime-filling threebranes 
for tadpole cancellation [\ref{SVW}]. This should be compared with a
potentially dual heterotic compactification on an elliptic Calabi-Yau $Z$
with vector bundle $V$ embedded in $E_8\times E_8$ got by adiabtic extension 
over a common (complex) twofold base $B$ of
the eight-dimensional duality of $F$-theory on $K3$ with
the heterotic string on $T^2$ [\ref{V}].
There the threebranes should correspond to a number $n_5$ of 
fivebranes wrapping the elliptic
fibre. Their necessity for a consistent heterotic compactification
(independent of any duality considerations) was established in the 
exhaustive study done by Friedman, Morgan and Witten on vector budles and
$F$-theory [\ref{FMW}]. There it was also shown that in the case of an
$E_8\times E_8$ vector bundle $V$, leaving no unbroken gauge group and
corresponding to a smooth Weierstrass model for the fourfold, it is possible
to express $n_3$ and $n_5$ in comparable and indeed matching data on the 
common base $B$. This matching will here be extended to the general case.

 To achieve this we will adopt a somewhat different technical 
procedure.
In the case of not having a smooth Weierstrass model there occured already
in [\ref{SVW}] the difficulty to reduce the fourfold expression
$\chi (X)/24$ for $n_3$ to an expression involving only suitable
data of the base $B^3$ of the elliptic $F$-theory fibration of $X$ (not to
be confused with the twofold base $B^2$ of the $K3$ fibration of $X$, here
denoted simply by $B$, which is visible also on the heterotic side),
which was [\ref{FMW}] only an intermediate step to reduce the expression
to one involving only twofold base data.
For this reason we express here $\chi (X)$ directly in the Hodge numbers 
of $X$ and match then these with the data of the dual heterotic model;
here essential use is made of an index-formula computation, also initiated in
[\ref{FMW}], to describe the moduli of the bundle $V$.\footnote{Note that in 
case some of the threebranes have dissolved into finite-sized
instantons in the world-volume gauge theory of the $F$-theory sevenbrane
this is accompanied by a corresponding fivebrane transition on the heterotic
side [\ref{BJPS}, \ref{FMW}, \ref{S}].}

 In the course of the actual computation we will assume that $V$ is a
$SU(n_1)\times SU(n_2)$ bundle. Furthermore we will implement the adiabatically
extended duality by the following specification [\ref{FMW}]: as $X$ is assumed
to be a $K3$ fibration over $B$ it follows that $B^3$, the threefold base 
of the $F$-theory elliptic fibration, is a $P^1$ fibration over $B$; this 
fibration structure is described by assuming the $P^1$ bundle over $B$
to be a projectivization of a vector bundle $Y={\cal O}\oplus {\cal T}$, with
${\cal T}$ a line bundle over $B$; then the cohomology class $t=c_1({\cal T})$
encodes the $P^1$ fibration structure. Now the duality is implemented by
choosing our $SU(n_1)\times SU(n_2)$ bundle so that

\beqa
\eta_1=\pi_*(c_2(V_1))=6c_1(B)+t, \ \ \  \eta_2=\pi_*(c_2(V_2))=6c_1(B)-t
\nonumber
\eeqa

 Now let us start and first compute the number of heterotic five-branes 
$n_5$ using the 
non-perturbative anomaly cancellation condition which is given by
[\ref{FMW}]

\beqa
c_2(V_1)+c_2(V_2)+[W]=c_2(TZ),
\nonumber
\eeqa 
where $TZ$ is tangent bundle of the Calabi-Yau threefold $Z$ on which
the heterotic string is compactified; $c_2(TZ)$ was derived in [\ref{FMW}]
(we use the notation $c_i=c_i(B)$ and $\sigma$ for the class of a 
section of $\pi$)
\beqa
c_2(TZ)=c_2+11c_1^2+12\sigma c_1\nonumber
\eeqa
(here on the right hand side the classes represent their pullbacks to $Z$).
Furthermore $[W]$ is the cohomology class 
of the five-branes where $[W]=n_5[F]=\pi^*([p])$ and $[F]$ is the class 
in $H^4$ of the fiber of the elliptic fibration.

For explicit evaluation and to establish some notation we give here the 
actual number of fivebranes for $V$ a $SU(n_1)\times SU(n_2)$ bundle.
The second Chern class for a $SU(n)$ bundle
is given by (with $\eta= \pi_*(c_2(V))$ and ${\cal L}$ being some line
bundle over $B$) [\ref{FMW}]

\beqa
c_2(V)= \eta\sigma -\frac{1}{24}c_1({\cal L})^2(n^3-n)-\frac{n}{8}\eta(
                                      \eta-nc_1({\cal L}))
\nonumber
\eeqa

Using the above relations
for $\eta_1$, $\eta_2$ and $c_2(V)$ we can derive 
(where also $c_1(B)=c_1({\cal L})$ by the 
Calabi-Yau condition for $Z$)

\beqa
c_2(V_{1/2})=6c_1\sigma \pm t\sigma - \frac{1}{24}c_1^2
            (n_1^3-n_1)- \frac{n_1}{8}[36c_1^2 \pm 12c_1\, t
            +t^2 -6n_1c_1^2\mp n_1\, t\, c_1]
\nonumber
\eeqa  

We find for the number of five branes

\beqa
n_5&=& c_2+11c_1^2+\frac{1}{24}c_1^2(n_1^3-n_1+n_2^3-n_2)+
       \frac{(n_1+n_2)}{4}(18c_1^2+\frac{t^2}{2})+\nonumber\\
   & & +\frac{(n_1-n_2)}{4}6c_1\, t-
       \frac{(n_1^2+n_2^2)}{8}
       6c_1^2+\frac{(n_2^2-n_1^2)}{8}\, t\, c_1
\nonumber
\eeqa

 Now let us express the number of $F$-theory 3-branes
$n_3=\chi (X)/24$ in terms of heterotic data. Because of
$\chi (X)/6 -8=h^{1,1}(X)+h^{3,1}(X)-h^{2,1}(X)$ (cf. [\ref{SVW}])
we have to use the following informations \footnote{
We assume that the elliptic fibration $Z\rightarrow B$ has only
one section so that $h^{1,1}(Z)=h^{1,1}(B)+1=c_2(B)-1=11-c_1^2(B)$;
let us furthermore assume that no fourflux [\ref{W4fl}] is turned on;
$r$ denotes the rank of the unbroken non-abelian gauge group (do not confuse
it with the rank of the group of the bundle $V$); we furthermore assume that 
we have no further $U(1)$ factors (coming from sections).}

\beqa
h^{1,1}(X)&=& h^{1,1}(Z)+1+r=12-c_1^2+r\nonumber\\
h^{2,1}(X)&=& n_o\nonumber\\
h^{3,1}(X)&=& h^{2,1}(Z)+I+n_o+1=12+29 c_1^2+I+h^{2,1}(X)\nonumber
\eeqa
where in the first line
we used Noethers formula $1=\frac{c_1^2(B)+c_2(B)}{12}$ and
$\chi (Z)=-60\int _B c_1^2(B)$ [\ref{KLRY}]; for the $n_o$ in the second line
see below. Let us now see how the 
expression for $h^{3,1}(X)$ emerges.
Note that
the moduli space ${\cal M}$ for bundles on $Z$ of dimension $m_{bun}$ 
has a fibration ${\cal M}\rightarrow {\cal Y}$ which corresponds on the 
$F$-theory side to a fibration of the abelian variety 
$H^3(X,{\bf R})/H^3(X,{\bf Z})$ of complex dimension $h^{2,1}(X)$ over
a part of the space of complex deformations of $X$ (cf. [\ref{FMW}]); the 
remaining complex deformations account for the complex deformations of 
the heterotic Calabi-Yau (+1), i.e.

\beqa
h^{3,1}(X)+h^{2,1}(X)=h^{2,1}(Z)+m_{bun}+1\nonumber
\eeqa

 Now concerning the contribution of the bundle moduli let us recall the
setup of the index-computation in [\ref{FMW}]. As the usual quantity
suitable for index-computation $\sum_{i=o}^3 (-1)^i h^i(Z,ad(V))$ vanishes
by Serre duality one has to introduce a further twist and to compute a
character-valued index. Now because of the elliptic fibration structure
one has on $Z$ the involution $\tau$ coming from the "sign-flip" in the fibers
which we furthermore assume has been lifted to an action on the bundle.
The character-valued index

\begin{eqnarray}
I=-\frac{1}{2}\sum_{i=o}^3 (-1)^i Tr_{H^i(Z,ad(V))} \tau \nonumber
\end{eqnarray}
simplifies by the vanishing of the ordinary index to

\begin{eqnarray}
I=-\sum_{i=o}^3 (-1)^i h^i(Z,ad(V))_e \nonumber
\end{eqnarray}
where the subscript "e" (resp. "o") indicates the even (resp. odd) part. As
we have an unbroken gauge group $H$, which is the commutator of the group
$G$ of $V$, one finds $I=n_e-n_o$ corrected by $h^0_e-h^0_o$
denoting by $n_{e/o}$  the number $h^1(Z,ad(V))_{e/o}$ of massless even/odd
chiral superfields and by $h^0_{e/o}$ the number of unbroken gauge group
generators even/odd under $\tau$. The unbroken gauge group is in this $n_3$-
calculation accounted for by the rank contribution in $h^{1,1}(X)$ for the 
resolved fourfold.

So one has for the number of bundle moduli $m_{bun}=h^1(Z,ad(V))=
n_e+n_o=I+2n_o$ and so
gets the announced expression for $h^{3,1}(X)$. Furthermore on the 
$F$-theory side the modes odd under the involution $\tau ^{\prime}$ 
corresponding to the heterotic involution $\tau$ correspond to the 
$h^{2,1}(X)$ classes [\ref{FMW}] (we assume no four-flux was turned on).

Now let us go to the index-formula for a $SU(n_1)\times SU(n_2)$
bundle. The index for $SU(n)$ is given by [\ref{FMW}]

\beqa
I=n-1-\int_{U_{1}}c_{2}(V)-\int_{U_{2}}c_2(V).
\nonumber
\eeqa
Recall the $U_{i}$ are the components of the fixed point set 
of the involution $\tau$ where 
$U_1$ is the class of the section $\sigma$ of $Z\rightarrow B$ and
$U_{2}$ is a triple cover of $B$ which has cohomology class 
$3\sigma+3c_1({\cal L})$. Let us go on and express this purely in 
cohomology of $B$. This can be done in two ways. First one can use the
identity $\int_{U_{i}}c_2(V)=\int_{B}c_2(V)\mid_{U_{1}}+3\int_{B}c_2(V)
\mid_{U_{2}}$ where the 3 appears since $U_{2}$ is a triple cover of 
$B$ and restricts the non-perturbative anomaly condition to the
fixed point set (cf. [\ref{ACL}]).
The restricted second Chern-classes of $TZ\mid_{U_{i}}$ 
were direved in [\ref{ACL}] as $c_2(TZ\mid_{U_{1}})=c_2-c_1^2$ and
$c_2(TZ\mid_{U_{2}})=c_2+11c_1^2$.
Thus for the index we find (denoting $rk=n_1+n_2-2=16-r)$)

\begin{eqnarray}
I=rk-(c_2-c_1^2-n_5+3c_2+33c_1^2-3n_5)=rk-(48+28c_1^2)+4n_5\nonumber
\end{eqnarray} 

 Alternatively one can use formula for the 
restriction of $c_2(V)$ to the fixed point set
given in [\ref{FMW}]

\beqa
I&=&n-1+\int_{B}\Bigg(  \frac{(n^3-n)c_1({\cal L})^2}{6}+\frac{n\eta(\eta-
             nc_1({\cal L}))}{2}+\eta c_1({\cal L}) \Bigg)\nonumber\\
 &=&n-1-4(c_2-\eta \sigma)+\eta c_1\nonumber
\eeqa
giving again for the total index $I=I_1+I_2$ the result

\beqa
I&=&rk-4(c_2(V_1)+c_2(V_2))+48c_1 \sigma +12c_1^2\nonumber\\
 &=& rk-4(c_2+11c_1^2+12\sigma c_1)+4n_5+
     48c_1 \sigma +12c_1^2\nonumber\\
 &=&rk-(48+28c_1^2)+4n_5\nonumber
\eeqa
One finally gets then with $r=16-rk$ 
and our expression for $I$ that

\beqa
\chi (X)/24=2+\frac{1}{4}(12-c_1^2+16-rk+12+29 c_1^2+I)=n_5\nonumber
\eeqa
matching the heterotic value.

\section*{References}
\begin{enumerate}

\item
\label{V}
C. Vafa, {\it Evidence for F-theory}, Nucl. Phys. {\bf B 469} (1996) 493,
hep-th/9602022.

\item
\label{FMW}
R. Friedman, J. Morgan and E. Witten, {\it Vector Bundles and 
F-Theory}, hep-th/9701162.

\item
\label{SVW}
S. Sethi, C. Vafa and E. Witten, {\it Constraints on Low-Dimensional String
Compactifications}, Nucl. Phys. {\bf B 480} (1996) 213, hep-th/9606122.

\item
\label{S}
E. Sharpe, {\em Extremal Transitions in Heterotic String Theory}, 
hep-th/9705210.

\item
\label{KLRY}
A. Klemm, B. Lian, S.-S. Roan and S.-T. Yau, {\it Calabi-Yau fourfolds
for $M$- and $F$-Theory Compactifications}, hep-th/9701023.

\item
\label{W4fl}
E. Witten, {\it On Flux Quantization in M-Theory and the Effective Action},
hep-th/9609122.

\item
\label{ACL}
B. Andreas, G. Curio and D. L\"ust, {\it $N=1$ Dual String Pairs and their
Massless Spectra}, hep-th/9705174.

\item
\label{BJPS}
M. Bershadsky, A. Johansen, T. Pantev and V. Sadov, {\it On Four-Dimensional 
Compactifications of F-Theory}, hep-th/9701165.

\end{enumerate}

\end{document}